# *A community perspective on the concept of marine holobionts: state-of-the-art, challenges, and future directions*

The Holomarine working group[1]: Simon M. Dittami[†], Enrique Arboleda, Jean-Christophe Auguet, Arite Bigalke, Enora Briand, Paco Cárdenas, Ulisse Cardini, Johan Decelle, Aschwin Engelen, Damien Eveillard, Claire M.M. Gachon, Sarah Griffiths, Tilmann Harder, Ehsan Kayal, Elena Kazamia, Francois H. Lallier, Mónica Medina, Ezequiel M. Marzinelli, Teresa Morganti, Laura Núñez Pons, Soizic Pardo, José Pintado Valverde, Mahasweta Saha, Marc-André Selosse, Derek Skillings, Willem Stock, Shinichi Sunagawa, Eve Toulza, Alexey Vorobev, Catherine Leblanc[†], and Fabrice Not[†]

[†] Corresponding authors: Simon M Dittami (simon.dittami@sb-roscoff.fr), Catherine Leblanc (catherine.leblanc@sb-rosocff.fr), and Fabrice Not (fabrice.not@sb-roscoff.fr)

Simon M. Dittami, simon.dittami@sb-roscoff.fr, Sorbonne Université, CNRS, Integrative Biology of Marine Models (LBI2M), Station Biologique de Roscoff, 29680 Roscoff, France

Enrique Arboleda, arboleda.enrique@gmail.com, Sorbonne Université, CNRS, FR2424, Station Biologique de Roscoff, 29680 Roscoff, France

Jean-Christophe Auguet, jean-christophe.auguet@cnrs.fr, MARBEC, Université de Montpellier, CNRS, IFREMER, IRD, Montpellier, France

Arite Bigalke, arite.bigalke@uni-jena.de, Institute for Inorganic and Analytical Chemistry, Bioorganic Analytics, Friedrich-Schiller-Universität Jena, Lessingstrasse 8, D-07743 Jena, Germany

Enora Briand, enora.briand@ifremer.fr, Ifremer, Laboratoire Phycotoxines, 44311 Nantes, France

Paco Cárdenas, paco.cardenas@ilk.uu.se, Pharmacognosy, Department of Medicinal Chemistry, Uppsala University, BMC Box 574, 75123 Uppsala, Sweden

Ulisse Cardini, ulisse.cardini@szn.it, Integrative Marine Ecology Department, Stazione Zoologica Anton Dohrn, Napoli, Italy

Johan Decelle, johan.decelle@univ-grenoble-alpes.fr, Laboratoire de Physiologie Cellulaire et Végétale, Université Grenoble Alpes, CNRS, CEA, INRA; 38054, Grenoble Cedex 9, France.

---

[1] This working group gathers 31 scientists from ten different countries, with expertise covering different scientific disciplines including philosophy, evolution, computer sciences, marine biology, ecology, chemistry, and microbiology, who participated in a workshop on marine holobionts, organized at the Roscoff Biological Station in March 2018. Their aim was to exchange ideas regarding key concepts and opportunities in marine holobiont research, to start structuring the community, and to identify and tackle key challenges in the field.

Aschwin Engelen, aengelen@ualg.pt, CCMAR, Universidade do Algarve, Campus de Gambelas, Faro, Portugal

Damien Eveillard, damien.eveillard@univ-nantes.fr, Université de Nantes, CNRS, Laboratoire des Sciences Numériques de Nantes (LS2N), 44322 Nantes, France

Claire M.M. Gachon, claire.gachon@sams.ac.uk, Scottish Association for Marine Science Scottish Marine Institute, PA37 1QA Oban, United Kingdom

Sarah Griffiths, griffiths.sarahm@gmail.com, School of Science and the Environment, Manchester Metropolitan University, Manchester, UK

Tilmann Harder, t.harder@uni-bremen.de, University of Bremen, Leobener Strasse 6, 28359 Bremen, Germany

Ehsan Kayal, ehsan.kayal@sb-roscoff.fr, Sorbonne Université, CNRS, FR2424, Station Biologique de Roscoff, 29680 Roscoff, France

Elena Kazamia, kazamia@biologie.ens.fr, Institut de Biologie de l'Ecole Normale Supérieure, 46 rue d'Ulm, 75005 Paris, France

François H. Lallier, lallier@sb-roscoff.fr, Sorbonne Université, CNRS, Adaptation and Diversity in the Marine Environment, Station Biologique de Roscoff, 29680 Roscoff, France

Mónica Medina, mum55@psu.edu Department of Biology, Pennsylvania State University, University Park PA 16801, USA

Ezequiel M. Marzinelli, e.marzinelli@sydney.edu.au, 1) The University of Sydney, School of Life and Environmental Sciences, Sydney, NSW 2006, Australia; 2) Singapore Centre for Environmental Life Sciences Engineering, Nanyang Technological University, Singapore; 3) Sydney Institute of Marine Science, Mosman, NSW 2088, Australia

Teresa Morganti, tmorgant@mpi-bremen.de, Max Planck Institute for Marine Microbiology, Celsiusstraße 1, 28359, Bremen, Germany

Laura Núñez Pons, laura.nunezpons@szn.it, Section Biology and Evolution of Marine Organisms (BEOM), Stazione Zoologica Anton Dohrn (SZN), Villa Comunale, 80121, Napoli, Italy

Soizic Prado, sprado@mnhn.fr, Molecules of Communication and Adaptation of Microorganisms (UMR 7245), National Museum of Natural History, CNRS, CP 54, 57 rue Cuvier, 75005 Paris, France

José Pintado, pintado@iim.csic.es, Instituto de Investigaciones Marinas (IIM-CSIC), Eduardo Cabello 6, 36208 Vigo, Galicia, Spain

Mahasweta Saha, sahamahasweta@gmail.com, 1) Benthic Ecology, Helmholtz Center for Ocean Research, Düsternbrooker Weg 20, 24105 Kiel, Germany; 2) School of Biological Sciences, University of Essex, Wivenhoe Park, CO4 3SQ, Colchester, United Kingdom.

Marc-André Selosse, ma.selosse@wanadoo.fr, 1) Département Systématique et Evolution, Muséum national d'Histoire naturelle, UMR 7205 ISYEB, CP 50, 45 rue Buffon, Paris 75005, France ; 2) Faculty of Biology, University of Gdansk, ul. Wita Stwosza 59, 80- 308, Gdansk, Poland

Derek Skillings, derek.skillings@gmail.com, Philosophy Department, University of Pennsylvania, 249 S. 36th Street, Philadelphia PA 19104-6304, USA

Willem Stock, Willem.Stock@ugent.be, Laboratory of Protistology & Aquatic Ecology, Department of Biology, Ghent University, Krijgslaan 281-S8, 9000 Ghent, Belgium

Shinichi Sunagawa, ssunagawa@ethz.ch, Department of Biology, Institute of Microbiology and Swiss Institute of Bioinformatics, ETH Zürich, Vladimir-Prelog-Weg 4, 8093 Zürich, Switzerland

Eve Toulza, eve.toulza@univ-perp.fr, Univ. Perpignan Via Domitia, IHPE UMR 5244, CNRS, IFREMER, Univ. Montpellier, 66000 Perpignan, France

Alexey Vorobev, voralexey@gmail.com, CEA - Institut de Biologie François Jacob, Genoscope, 2 Rue Gaston Crémieux, 91057 Evry, France

Catherine Leblanc, catherine.leblanc@sb-roscoff.fr, Sorbonne Université, CNRS, Integrative Biology of Marine Models (LBI2M), Station Biologique de Roscoff, 29680 Roscoff, France

Fabrice Not, fabrice.not@sb-roscoff.fr, Sorbonne Université, CNRS, Adaptation and Diversity in the Marine Environment (AD2M), Station Biologique de Roscoff, 29680 Roscoff, France

## Abstract:

Host-microbe interactions play crucial roles in marine ecosystems, but we still have very little understanding of the mechanisms that govern these relationships, the evolutionary processes that shape them, and their ecological consequences. The holobiont concept is a renewed paradigm in biology that can help describe and understand these complex systems. It posits that a host and its associated microbiota, living together in a long-lasting relationship, form the holobiont, and have to be studied together, as a coherent biological and functional unit, in order to understand the biology, ecology and evolution of the organisms. Here we discuss critical concepts and opportunities in marine holobiont research and identify key challenges in the field. We highlight the potential economic, sociological, and environmental impacts of the holobiont concept in marine biological, evolutionary, and environmental sciences with comparisons to terrestrial science whenever appropriate. A deeper understanding of such complex systems, however, will require further technological and conceptual advances. The most significant challenge will be to bridge functional research on simple and tractable model systems and global approaches. This will require scientists to work together as an (inter)active community in order to address, for instance, ecological and evolutionary questions and the roles of holobionts in biogeochemical cycles.

## Glossary

**Anna Karenina principle** – a number of factors can cause a system to fail, but only a narrow range of parameters characterizes a working system; based on the first sentence of Leo Tolstoy's "Anna Karenina": "Happy families are all alike; every unhappy family is unhappy in its own way."

**Dysbiosis** – microbial imbalance in a symbiotic community that affects the health of the host.
**Ecosystem services** – any direct or indirect benefits that humans can draw from an ecosystem; they include provisioning services (*e.g.*, food), regulating services (*e.g.*, climate), cultural services (*e.g.*, recreation), and supporting services (*e.g.*, habitat formation).
**Endosymbiosis** – a symbiotic relationship in which a symbiont lives inside a host; a prominent example are plant root nodules.
**Gnotobiosis** – the condition in which all organisms present in a culture can be controlled.
**Holobiont** – an ecological unit of different species living together in a long-lasting relationship.
**Horizontal transmission** – acquisition of the associated microbiome by a new generation of hosts from the environment.
**Host** – the largest partner in a symbiotic community.
**Infochemical** – a diffusible chemical compound used to exchange information between organisms.
**Microbial gardening** – behavior, frequently the release of growth-enhancing or inhibiting chemicals or metabolites that favors the development of a microbial community beneficial to the host.
**Microbiome** – the combined genetic information encoded by the microbiota; may also refer to the microbiota itself.
**Microbiota** – all microorganisms present in a particular environment or associated with a particular host.
**Nested ecosystems** – a view of ecosystems where each individual system can be decomposed into smaller systems and/or considered part of a larger system, all of which still qualify as ecosystems.
**Phagocytosis** – a process by which a eukaryotic cell ingests other cells or solid particles.
**Phylosymbiosis** – congruence in the phylogeny of different hosts and the composition of their associated microbiota.
**Rasputin effect** – the phenomenon that commensals and mutualists can become parasitic in certain conditions; after the Russian monk Rasputin who became the confidant of the Tsar of Russia, but later helped bring down the Tsar's empire during the Russian revolution.
**Sponge loop** – sponges efficiently recycle dissolved organic matter turning it into detritus that becomes food for other consumers.
**Symbiont** – an organism living in symbiosis; usually refers to the smaller/microbial partners living in commensalistic or mutualistic relationships (see also host).
**Symbiosis** – a close and long-lasting relationship between organisms living together; includes mutualistic, commensalistic, and parasitic relationships.
**Vertical transmission** – acquisition of the associated microbiome by a new generation of hosts from the parents (contrary to horizontal transmission).

# Marine holobionts from their origins to the present

## The history of marine holobiont concept

Current theory proposes a single origin for eukaryotic cells through the symbiotic assimilation of prokaryotes to form cellular organelles such as plastids and mitochondria (reviewed in Archibald 2015). These ancestral and founding symbiotic events, which prompted the metabolic and cellular complexity of eukaryotic life, most likely occurred in the ocean, where eukaryotic phagocytosis s widespread (Martin *et al.* 2008).

Despite the general acceptance of this so-called endosymbiotic theory, the term 'holobiont' did not immediately enter the scientific vernacular. It was coined by Lynn Margulis in 1990, who proposed that evolution has worked mainly through symbiosis-driven leaps that merged organisms into new forms referred to as 'holobionts', and only secondarily through gradual mutational changes (Margulis 1990; O'Malley 2017). However, the concept did not become widely used until it was co-opted by coral biologists over a decade later. Corals and dinoflagellate algae of the family Symbiodiniaceae are one of the most iconic examples of symbioses found in nature; most corals are incapable of long-term survival without the products of photosynthesis provided by their algae. Rohwer *et al.* (2002) were the first to use the word "holobiont" to describe corals, where the holobiont comprised the coral organism (host), Symbiodiniaceae, endolithic algae, prokaryotes, fungi, other unicellular eukaryotes, and viruses, together used to describe a unit of selection *sensu* Margulis (Rosenberg *et al.* 2007b).

Although initially driven by studies of marine organisms, much of the research on the emerging properties and significance of holobionts has since been carried out in other fields of research: the microbiota of the rhizosphere of plants or the animal gut became predominant models and have led to an ongoing paradigm change in agronomy and medical sciences (Bulgarelli *et al.* 2013; Shreiner *et al.* 2015; Faure *et al.* 2018). Holobionts occur in all habitats, terrestrial and aquatic, and several analogies between these ecosystems can be made. For example, it is clear that interactions within holobionts are mediated by chemical cues in the environment, so-called infochemicals; (Loh *et al.* 2002; Rolland *et al.* 2016; Saha *et al.* 2019). The major differences across systems are notably due to the physical nature of water, which often ensures stronger physicochemical interconnections among niches and habitats. In marine ecosystems, carbon fluxes also appear to be swifter and trophic modes more flexible, leading to higher plasticity of functional interactions (Mitra *et al.* 2013). Moreover, dispersal barriers are usually lower, allowing for faster microbial shifts in marine holobionts (Kinlan and Gaines 2003; Martin-Platero *et al.* 2018). Finally, phylogenetic diversity at broad taxonomic scales (*i.e.*, suprakingdom, kingdom, phyla levels), is higher in the sea than on land, with much of the marine diversity yet to be uncovered (de Vargas *et al.* 2015; Thompson *et al.* 2017), notably for marine viruses (Middelboe and Brussaard 2017). The recent discovery of this astonishing marine microbial diversity and the scarcity of marine holobiont research suggest a high potential for complex cross lineage interactions yet to be explored in marine holobiont systems (Figure 1).

# Evolution of holobionts

These examples and the associated debate over how to define organisms or functional entities led to the revival of 'holism', the philosophical notion, first proposed by Aristotle. Since the Age of Enlightenment and the shift toward "dissection science", one dominant thought in sciences was to focus on the smallest component of a system to understand it better. Holistic thinking states that systems should be studied in their entirety, with a focus on the interconnections between the individual parts rather than the parts themselves (Met. Z.17, 1041b11–33). Such systems have emergent properties that result from the irreducible behavior of a system that is 'larger than the sum of its parts'. The boundaries of holobionts are usually delimited by a physical envelope, which corresponds to the area of local influence of the host. However, they may also be defined in a context-dependent way as a 'Russian Matryoshka doll' encompassing all the levels of host-microbiota associations up to the community and ecosystem level, a concept referred to as "nested ecosystems" (Figure 2; McFall-Ngai *et al.* 2013; Pita *et al.* 2018). Such a view raises fundamental questions for studies of evolution, regarding the relevant units of selection and the role of co-evolution. For instance, plant and animal evolution involves new functions co-constructed by members of the holobiont or elimination of functions redundant between them (Selosse *et al.* 2014). Rosenberg and Zilber-Rosenberg (2018) have argued that all animals and plants can be considered as holobionts, and thus advocated the hologenome theory of evolution. It proposes that natural selection acts at the level of the holobiont and the hologenome (*i.e.,* the combined genomes of the host and all members of its microbiota; Rosenberg *et al.* 2007a; Zilber-Rosenberg and Rosenberg 2008). This interpretation of Margulis' definition of a 'holobiont' considerably broadened fundamental concepts in evolution and speciation and has not been free of criticism (Douglas and Werren 2016). It is, however, generally recognized that, although it should not be accepted as a default for explaining features of host-symbiont associations (Moran and Sloan 2015), the holobiont needs to be at least considered when addressing evolutionary and ecological questions.

# Marine holobiont models

Today, an increasing number of marine organisms are being used as holobiont model systems in research, often with a different emphasis, but altogether covering a large range of scientific topics. Examples of this diversity and related insights are provided in this section. Spongebacteria interactions are particularly promising for the discovery of novel bacterial lineages or new drugs (Blunt *et al.* 2016; Bibi *et al.* 2017). The flatworm *Symsagittifera* (= *Convoluta*) *roscoffensis* (Arboleda *et al.* 2018), the sea anemone *Exaiptasia* (Baumgarten *et al.* 2015; Wolfowicz *et al.* 2016), the upside-down jellyfish *Cassiopea* (Ohdera *et al.* 2018), and their respective intracellular green and dinoflagellate algae have, in addition to corals, become models for fundamental research on evolution of metazoan-algal photosymbiosis. Similarly, radiolarians and foraminiferans (both heterotrophic protists dwellers harboring endosymbiotic microalgae) are emerging as critical ecological models for unicellular photosymbiosis due to their ubiquitous presence in the world's oceans (Decelle *et al.* 2015; Not *et al.* 2016). The discovery of deep-sea hydrothermal vents revealed symbioses of animals with chemosynthetic bacteria that have later been found in many other marine ecosystems (Dubilier *et al.* 2008; Rubin-Blum *et al.* 2019) and frequently exhibited high levels of metabolic and taxonomic diversity (Duperron *et al.* 2008; Petersen *et al.* 2016; Ponnudurai *et al.* 2017).The *Vibrio*-squid model provides insights into the effect of microbiotas on animal development, circadian rhythms, and immune systems (McFallNgai 2014). The cosmopolitan haptophyte *Emiliania huxleyi*, promoted by associated bacteria (Seyedsayamdost *et al.* 2011; Segev *et al.* 2016), produces important intermediates in the carbon and sulfur biogeochemical cycles making it an important model phytoplankton species. The green alga *Ostreococcus,* also an important player in marine primary production, has been shown to exchange vitamins with its microbiota (Cooper *et al.* 2019). The recent sequencing of several host genomes and their associated microorganisms, as well as improved experimental protocols, have led to new insights in many of these model systems, yet most experiments are still carried out in environmental or "semi-controlled" conditions. Only a few models, covering a small part of the overall marine biodiversity, are currently being cultivated *ex-situ* and can be used in fully controlled experiments, where they can be cultured aposymbiotically (*i.e.*, without symbionts). This is the case *e.g.* for the green macroalga *Ulva mutabilis*, which has enabled the exploration of bacteria-mediated growth and morphogenesis including the identification of original chemical interactions in the holobiont (Wichard 2015; Kessler *et al.* 2018) or for zooxanthellate sea anemones of the genus *Exaiptasia*, which have been used to explore photobiology disruption and restauration of cnidarian symbioses (Lehnert *et al.* 2012). Although the culture conditions in these highly controlled model systems are less realistic, we believe that such systems are essential to gain elementary mechanistic understanding of the functioning of marine holobionts.

## Marine holobionts as drivers of ecological processes

Marine holobionts can act as dissemination vectors for geographically restricted microbial taxa. For instance, pelagic mollusks or vertebrates have a high capacity for dispersion (*e.g.,* against currents and through stratified water layers). It has been estimated that fish and marine mammals may enhance the original dispersion rate of their microbiota by a factor of 200 to 200,000 (Troussellier *et al.* 2017) and marine birds may even act as bio-vectors across ecosystem boundaries (Bouchard Marmen *et al.* 2017). This holobiont-driven dispersal of microbes can include non-native or invasive species as well as pathogens (Troussellier *et al.* 2017). A related ecological function of holobionts is their potential to sustain rare species. Hosts provide an environment that favors the growth of specific microbial communities different from the surrounding environment (including rare microbes). They may, for instance, provide a nutrient-rich niche in the otherwise nutrient-poor seawater (Smriga *et al.* 2010; Webster *et al.* 2010; Burke *et al.* 2011; Chiarello *et al.* 2018), and the interaction between host and microbiota can allow both partners to cross biotope boundaries (*e.g.,* Woyke 2006) and colonize extreme environments (Bang *et al.* 2018). Holobionts thus contribute to marine microbial diversity and possibly resilience in the context of environmental change (Troussellier *et al.* 2017).

Microbially regulated biological processes are important drivers of global biogeochemical cycles (Falkowski *et al.* 2008; Madsen 2011; Anantharaman *et al.* 2016). These cycles describe the diverse global biogeochemical cycling are still sparse. In the open ocean, it is estimated that symbioses with the cyanobacterium UCYN-A contribute ~20% to the total $N_2$ fixation (Thompson *et al.* 2012; Martínez-Pérez *et al.* 2016). In benthic systems, sponges and corals may support entire ecosystems *via* their involvement in nutrient cycling thanks to their microbial partners (Raina *et al.* 2009; Fiore *et al.* 2010; Cardini *et al.* 2015; Pita *et al.* 2018), functioning as sinks/sources of nutrients. In particular the "sponge loop" recycles dissolved organic matter and makes it available to higher trophic levels in the form of detritus (de Goeij *et al.* 2013; Rix *et al.* 2017). In coastal sediments, bivalves hosting methanogenic archaea have been shown to increase the benthic methane efflux by a factor of up to eight, potentially accounting for 9.5% of total methane emissions from the Baltic Sea (Bonaglia *et al.* 2017).

Such impressive metabolic versatility is accomplished because of the simultaneous occurrence of disparate biochemical machineries (*e.g.,* aerobic and anaerobic pathways) in the individual symbionts, providing new metabolic abilities to the holobiont, such as the synthesis of specific essential amino acids, photosynthesis, or chemosynthesis (Venn *et al.* 2008; Dubilier *et al.* 2008). These metabolic capabilities have the potential to extend the ecological niche of the holobiont as well as its resilience to climate and environmental changes (Berkelmans and van Oppen 2006; Gilbert *et al.* 2010; Dittami *et al.* 2016; Shapira 2016; Godoy *et al.* 2018). It is therefore paramount to include the holobiont concept in predictive models that investigate the consequences of human impacts on the marine realm and its biogeochemical cycles.

# Challenges and opportunities in marine holobiont research

## Deciphering marine holobiont functioning

Two critical challenges that can be partially addressed by using model systems are 1) to decipher the factors determining holobiont composition and 2) to elucidate the impacts and roles of the different partners in these complex systems over time. Some marine invertebrates such as clams transmit part of the core microbiota maternally (Bright and Bulgheresi 2010; Funkhouser and Bordenstein 2013). In other marine holobionts, however, vertical transmission may be weak and inconsistent and mixed modes of transmission (vertical and horizontal) or intermediate modes (pseudo-vertical, where horizontal acquisition frequently involves symbionts of parental origin) are the most common (Bjork *et al.* 2018, preprint). Better understanding the factors that shape the composition of holobionts is highly relevant for marine organisms given that, despite a highly connected and microbe-rich environment, most marine hosts display a high specificity for their microbiota and even patterns of phylosymbiosis for some associations (Kazamia *et al.* 2016; Brooks *et al.* 2016; Pollock *et al.* 2018).

The immune system of the host is one way to regulate the microbial composition of the holobiont. Perturbations in this system can lead to dysbiosis, and eventually microbial infections (Selosse *et al.* 2014; de Lorgeril *et al.* 2018). Dysbiotic individuals frequently display higher variability in their microbial community composition than healthy individuals, an observation in line with the "Anna Karenina principle" (Zaneveld *et al.* 2017), although there are exceptions to this rule (*e.g.*, Marzinelli *et al.* 2015). A specific case of dysbiosis is the so-called "Rasputin effect" when benign endosymbionts opportunistically become detrimental to the host due to processes such as reduction in immune response under food deprivation, coinfections, or environmental pressure (Overstreet and Lotz 2016). Many diseases are now interpreted as the result of a microbial imbalance and the rise of opportunistic or polymicrobial infections upon host stress (Egan and Gardiner 2016). In reef-building corals, for example, warming destabilizes cnidarian-dinoflagellate associations, and some beneficial *Symbiodiniacea* strains switch their physiology and sequester more resources for their own growth at the expense of the coral host (Baker *et al.* 2018).

Another factor regulating holobiont composition is chemically mediated microbial gardening. This concept has already been demonstrated for land plants, where root exudates are used by plants to manipulate microbiome composition (Lebeis *et al.* 2015). In marine environments, comparable studies are only starting to emerge. For instance, seaweeds can chemically garden beneficial microbes aiding normal morphogenesis *via* exuded metabolites (Kessler *et al.* 2018), and corals of the genera *Acropora* and *Platygyra* structure their surfaceassociated microbiome by producing chemo-attractants and anti-bacterial compounds (Ochsenkühn *et al.* 2018). In the context of ongoing global change, an understanding of how the community and functional structure of resident microbes are resilient to perturbations remains critical to predict and assure the health of their host and the ecosystem.

## Integrating marine model systems with large-scale studies

By compiling what scientists today consider the most important trends and challenges in the field of marine holobiont research (Figure 3), we identified two distinct clusters: mechanistic understanding and predictive modeling. This illustrates that, on the one hand, the scientific community is focusing on the establishment of models for the identification of specific molecular interactions between marine organisms at a given point in space and time, up to the point of synthesizing mutualistic communities constituting functional marine holobionts *in vitro* (Kubo *et al.* 2013). On the other hand, another part of the community is moving towards global environmental sampling schemes such as the *TARA* Oceans expedition (Pesant *et al.* 2015) or the Ocean Sampling Day (Kopf *et al.* 2015), and towards long-term data series (*e.g.*, Wiltshire *et al.* 2010; Harris 2010). What emerges as both lines of research progress is the understanding that small-scale functional studies in the laboratory are inconsequential unless they are applicable to ecologically-relevant complex systems. At the same time, large scale-studies remain descriptive and with little predictive power unless we understand the mechanisms driving the observed processes. We illustrate the importance of integrating both approaches in Figure 3, where the node related to potential applications was perceived as a central hub at the interface between mechanistic understanding and predictive modeling.

A successful example allying both ends of the spectrum in terrestrial environments are the root nodules of legumes, which harbor nitrogen-fixing bacteria. In this system with a reduced number of symbionts involved, the functioning, distribution, and to some extent the evolution of these nodules, are now well understood (Epihov *et al.* 2017). The integration of this knowledge into agricultural practices has led to substantial yield improvements (*e.g.*, Kavimandan 1985; Alam *et al.* 2015). In the more diffuse and partner-rich system of mycorrhizal symbioses between plant roots and soil fungi, a better understanding of the interactions has also been achieved *via* the investigation of environmental diversity patterns in combination with experimental culture systems with reduced diversity (van der Heijden *et al.* 2015). We consider it essential to implement comparable efforts in marine sciences through interdisciplinary research combining biology, ecology, and mathematical modeling. Such approaches will enable the identification of common interaction patterns between organisms within holobionts and nested ecosystems. In addition to answering fundamental questions, they will help address the ecological, societal, and ethical issues that arise from attempting to actively manipulate holobionts (*e.g.*, in aquaculture) in order to enhance their resilience and protect them from the impacts of global change (Llewellyn *et al.* 2014).

# Emerging methodologies to approach the complexity of holobiont partnerships

As our conceptual understanding on the different levels of the organization of holobionts evolves, so does the need for multidisciplinary approaches and the development of tools and technologies to handle the unprecedented amount of data and their integration into dedicated models. Here progress is often fast-paced and provides exciting opportunities to address some of the challenges in holobiont research. Notably, a giant technological stride has been the explosion of affordable '–omics' technologies allowing molecular ecologists to move from metabarcoding (*i.e.*, sequencing of a taxonomic marker) to meta- or single-cell genomics, metatranscriptomics, and metaproteomics, thus advancing our understanding from phylogenetic to functional analyses of the holobiont (Bowers *et al.* 2017; Meng *et al.* 2018). For instance, metaproteomics combined with stable isotope fingerprinting can help study the metabolism of single species within the holobiont (Kleiner *et al.* 2018). In parallel, meta-metabolomics approaches have advanced over the last decades, and can be used to unravel the chemical interactions between partners. One current limitation here is that many compounds are still undescribed in databases and present in low quantities in natural environments, although recent technological advances such as molecular networking and meta-mass shift chemical profiling to identify relatives of known molecules promise significant improvements (Hartmann *et al.* 2017).

Additionally, it is highly challenging to identify the origin of a compound among the different partners of the holobionts and to determine its degree of involvement in the maintenance and performance of the holobiont. Well-designed experimental setups may help answer some of these questions (*e.g.,* Quinn *et al.* 2016), but they will also require high levels of replication due to extensive intra-species variability. Recently developed *in vivo* and *in situ* imaging techniques combined with 'omics' approaches can provide spatial and qualitative information (origin, distribution, and concentration of a molecule or nutrient), shedding new light on the role of each partner of the holobiont system at the subcellular level. The combination and stable isotope labelling and chemical imaging (mass spectrometry imaging such as secondary ion mass spectrometry and matrix-assisted laser desorption ionization, and synchrotron X-ray fluorescence), is particularly valuable in this context as it enables the investigation of metabolic exchange between the different components of a holobiont (Musat *et al.* 2016; Raina *et al.* 2017). Finally, three-dimensional electron microscopy may help evaluate to what extent different components of a holobiont are physically integrated (Colin *et al.* 2017; Decelle *et al.* 2019), where high integration is one indication of highly specific interactions.

   One consequence of the development of such new methods is the intellectual feedback they provide to improve existing models and to develop entirely new ones, for example by conceptualizing holobionts as mass balance of elements between the microbiota and its host (Skillings 2016; Berry and Loy 2018), or by redefining boundaries between the holobiont and the ecosystem (Zengler and Palsson 2012). Such models may incorporate metabolic complementarity between different components of the holobiont (Dittami *et al.* 2014; Bordron *et al.* 2016), simulate microbial communities starting from different cohorts of randomly generated microbes for comparison with actual metatranscriptomics and/or metagenomics data (Coles *et al.* 2017), or even employ machine learning techniques to predict host-associated microbial communities (Moitinho-Silva *et al.* 2017).

   A side-effect of these recent developments has been to shift the focus of holobiont research away from laboratory culture-based experiments. Here we argue that maintaining cultivation efforts to capture as much as possible of holobiont biodiversity remains essential in order to experimentally test hypotheses and investigate physiological mechanisms. A striking example of the importance of laboratory experimentation is the way germ-free mice reinoculated with cultivated bacteria (the so-called gnotobiotic mice) have contributed to the understanding of interactions within the holobiont in animal health and physiology (*e.g.,* Faith *et al.* 2014; Selosse *et al.* 2014). Innovations in cultivation techniques for axenic (or germ-free) hosts (*e.g.*, Spoerner *et al.* 2012) or in microbial cultivation such as microfluidic systems (*e.g.*, Pan *et al.* 2011) and cultivation chips (Nichols *et al.* 2010) may provide a way to obtain pure cultures. Yet, bringing individual components of holobionts into cultivation can still be a daunting challenge due to the strong interdependencies between organisms as well as the existence of yet unknown metabolic processes that may create specific requirements. In this context, single-cell omics analyses can provide critical information on some of the growth requirements of the organisms, and can complement approaches of high-throughput culturing (Gutleben *et al.* 2018). Established cultures can then be developed into model systems to move towards mechanistic understanding and experimental testing of hypothetical processes within the holobiont. A few such model systems have already been mentioned above, but omics techniques can broaden the range of available models, enabling generalizations about the functioning of marine holobionts and their interactions in marine environments (Wichard and Beemelmanns 2018).

## Ecosystem services and holobionts in natural and managed systems

A better understanding of marine holobionts can have straightforward socioeconomic consequences for coastal marine ecosystems, which have been estimated to provide services worth almost 50 trillion ($10^{12}$) US$ per year (Costanza *et al.* 2014). Most of the management practices of these systems have so far been based exclusively on the biology and ecology of macro-organisms. A multidisciplinary approach that provides mechanistic understanding of habitat-forming organisms as holobionts will ultimately improve the predictability and management of coastal ecosystems. For example, host-associated microbiotas could be integrated into the proxies used to assess the health of ecosystems. Microbial shifts and dysbiosis constitute early warning signals that may allow managers to predict potential impacts and intervene more rapidly and effectively (van Oppen *et al.* 2017; Marzinelli *et al.* 2018).

One form of intervention could be to promote positive changes of host-associated microbiotas, in analogous ways to the use of pre- and/or probiotics in humans (Singh *et al.* 2013) or inoculation of beneficial microbes in plant farming (Berruti *et al.* 2015; van der Heijden *et al.* 2015). In macroalgae, beneficial bacteria identified from healthy seaweed holobionts could be applied to diseased plantlets in order to suppress the growth of detrimental ones and/or to prevent disease outbreaks in aquaculture settings. In addition to bacteria, these macroalgae frequently host endophytic fungi that may have protective functions for the algae (Porras-Alfaro and Bayman 2011; Vallet *et al.* 2018). Host-associated microbiota could also be manipulated to shape key phenotypes in cultured marine organisms. For example, specific bacteria associated with microalgae may enhance their growth (Amin *et al.* 2009; Kazamia *et al.* 2012; Le Chevanton *et al.* 2013), increase their lipid content (Cho *et al.* 2015), and participate in the bioprocessing of algal biomass (Lenneman *et al.* 2014). More recently, the active modification of the coral microbiota has even been advocated as a means to boost the resilience of the holobiont to climate change (van Oppen *et al.* 2015; Peixoto *et al.* 2017), an approach which would, however, bear a high risk of unanticipated and unintended ecological consequences. Finally, one could implement holistic approaches in the framework of fish farms. Recent developments including integrated multi-trophic aquaculture, recirculating aquaculture, offshore aquaculture, and species selection and breeding increase yields and reduce the resource constraints and environmental impacts of intensive aquaculture (Klinger and Naylor 2012). However, the intensification of aquaculture often goes hand in hand with increased disease outbreaks both in industry and wild stocks. A holistic microbial management approach may provide an efficient solution to these latter problems (De Schryver and Vadstein 2014). Nevertheless, when considering their biotechnological potential, it should also be envisaged that marine microbiota may be vulnerable to anthropogenic influences and that their deliberate engineering, introduction from exotic regions, or inadvertent perturbations may have profound, and yet entirely unknown, consequences on marine ecosystems. Terrestrial environments provide examples of unwanted plant expansions or ecosystem perturbations linked to microbiota (*e.g.*, Dickie *et al.* 2017) and

PeerJ Preprints | NOT PEER-REVIEWED
examples where holobionts manipulated by human result in pests (*e.g.*, Clay and Holah 1999), calling for a cautious and ecologically-informed evaluation of holobiont-based technologies.

# Conclusions

Marine ecosystems represent highly connected reservoirs of largely unexplored biodiversity. They are of critical importance to feed the ever-growing world population, constitute a significant player in global biogeochemical cycles but are also threatened by human activities and global change. In order to unveil some of the basic principles of life and its evolution, and to protect and sustainably exploit marine natural resources, it is paramount to consider the complex biotic interactions that shape the marine communities and their environment. The scope of these interactions ranges from simple molecular signals between two partners *via* complex assemblies of eukaryotes, prokaryotes, and viruses with one or several hosts, to entire ecosystems. We believe that the concept of holobionts will be most useful and heuristic if used with a degree of malleability. It does not only represent the fundamental understanding that all living organisms have intimate connections with their immediate neighbors that may impact all aspects of their biology, but also enables us to define units of interacting organisms that are most suitable to answer specific scientific, societal, and economic questions. The holobiont concept marks a real paradigm shift in biological and environmental sciences, and a successful response to the underlying challenges will largely depend on the capacity of scientists to work together as an (inter)active community bringing together holistic and mechanistic views.

# Acknowledgments

This paper is based on the results of a foresight workshop funded by the EuroMarine network, Sorbonne University, and the UMRs 8227 and 7144 of the Roscoff Biological Station. We are grateful to Catherine Boyen for useful advice and helpful discussions. We thank Sylvie Kwayeb-Fagon for the workshop facilitation; Maryvonne Saout and Léna Corre for administrative support; and Marc Trousselier, Sébastien Villéger, Arthur Escalas, Yvan Bettarel, Thierry Bouvier for help writing a part of the manuscript. EMM was partially funded by an Australian Research Council Discovery Project (DP180104041), and JP was partially funded by the Galician Innovation Agency (IN607A 2017/4). The work of SD ad CL was partially funded by the ANR project IDEALG (ANR-10-BTBR-04). CG, CL, and SD received funding from the European Union's Horizon 2020 research and innovation programme under the Marie Sklodowska-Curie grant agreement number 624575 (ALFF). The work of FN was partially funded by the ANR project IMPEKAB (ANR-15-CE02-001). UC was partially funded by the Research Council of Lithuania project INBALANCE (09.3.3-LMT-K-712-01-0069). JD was supported by the LabEx GRAL (ANR-10-LABX-49-01) and Pôle CBS from the University of Grenoble Alpes. PC received support from the European Union's Horizon 2020 research and innovation program through the SponGES project (grant agreement No. 679849). EKAZ was funded by a Marie Curie Individual Fellowship (Horizon 2020, IRONCOMM). This document reflects only the authors' view and the Executive Agency for Small and Medium-sized Enterprises (EASME) is not responsible for any use that may be made of the information it contains.

# Figures

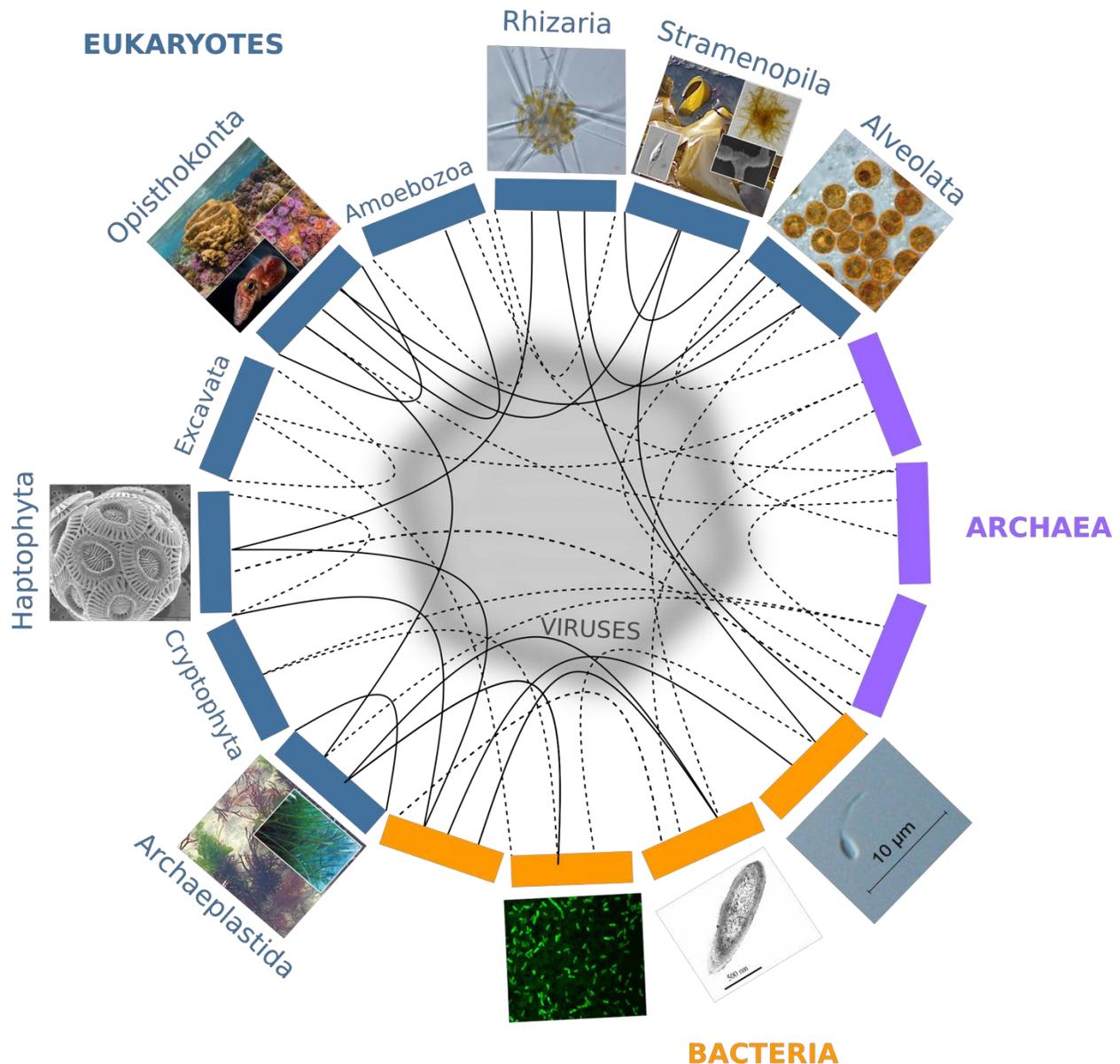

**Figure 1.** Partners forming marine holobionts are widespread across the tree of life including all kingdoms (eukaryotes, bacteria, archaea, viruses), and represent a large diversity of potential models for exploring complex biotic interactions across lineages. Plain lines correspond to holobionts referred to in the present manuscript. Dashed lines are examples of potential interactions. Photo credits: Archaeplastida - C. Leblanc, U Cardini; Stramenopila - C. Leblanc, S. M. Dittami, H. KleinJan; Alveolata - A. M. Lewis; Rhizaria - F. Not; Haptophyta - A. R. Taylor; Opisthonkonta - C. Frazee, M. McFall-Ngai, W. Thomas, L. Thiault; Bacteria - E Nelson, L Sycuro, S. M. Dittami, S. Le Panse.

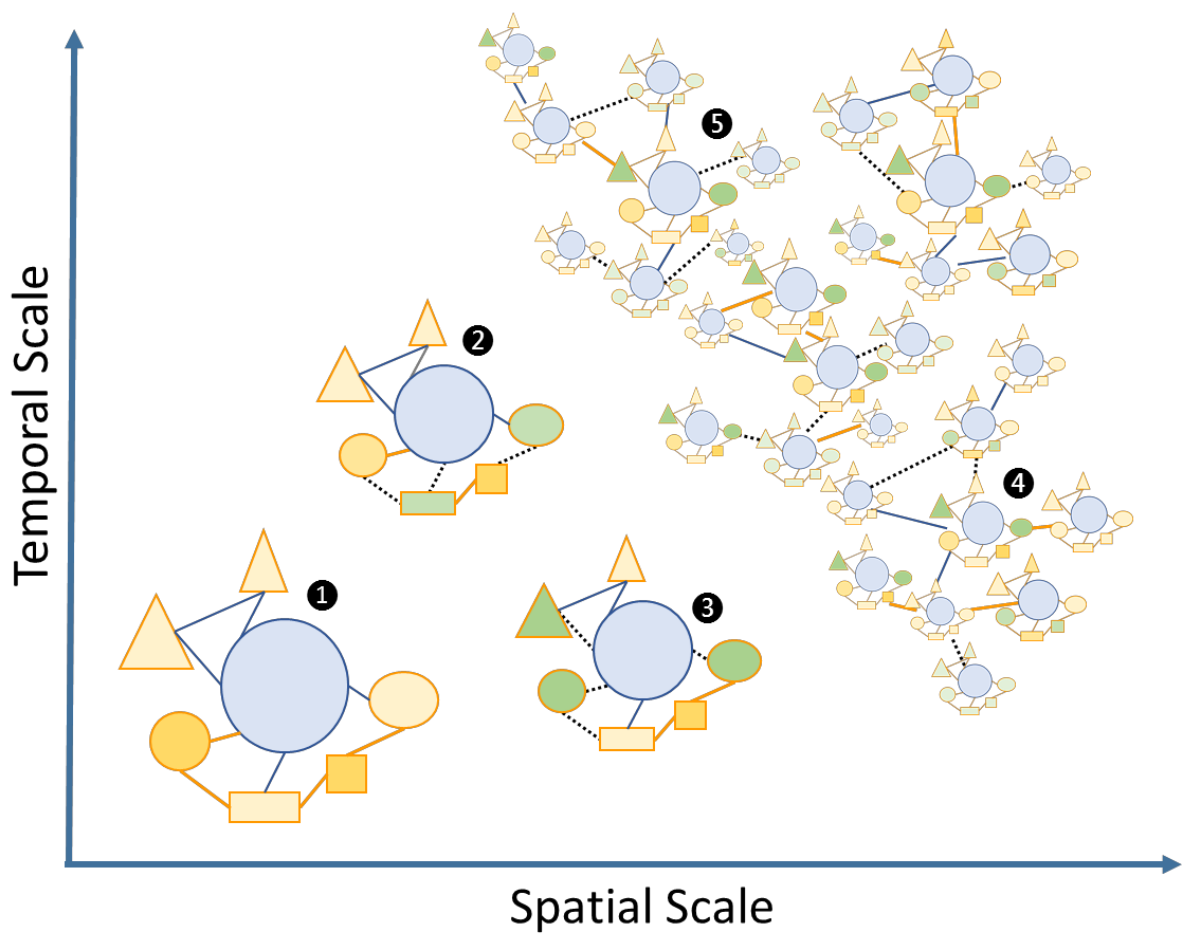

**Figure 2.** Schematic view of the "Russian Doll" complexity and dynamics of holobionts, according to diverse spatiotemporal scales. The host (blue circles), and associated microbes (all other shapes) including bacteria and eukaryotes that may be inside (*i.e.,* endosymbiotic or outside the host, are connected by either beneficial (solid orange lines), neutral (solid blue lines) or pathogenic (dashed black lines) interactions respectively. The different clusters can be illustrated by the following examples: 1, a model holobiont in one stable physiological condition (*e.g.*, in controlled laboratory condition); 2 and 3, holobionts changing during their life cycle or submitted to stress conditions; 4 and 5, marine holobionts in the context of global sampling campaign or long-term time series.

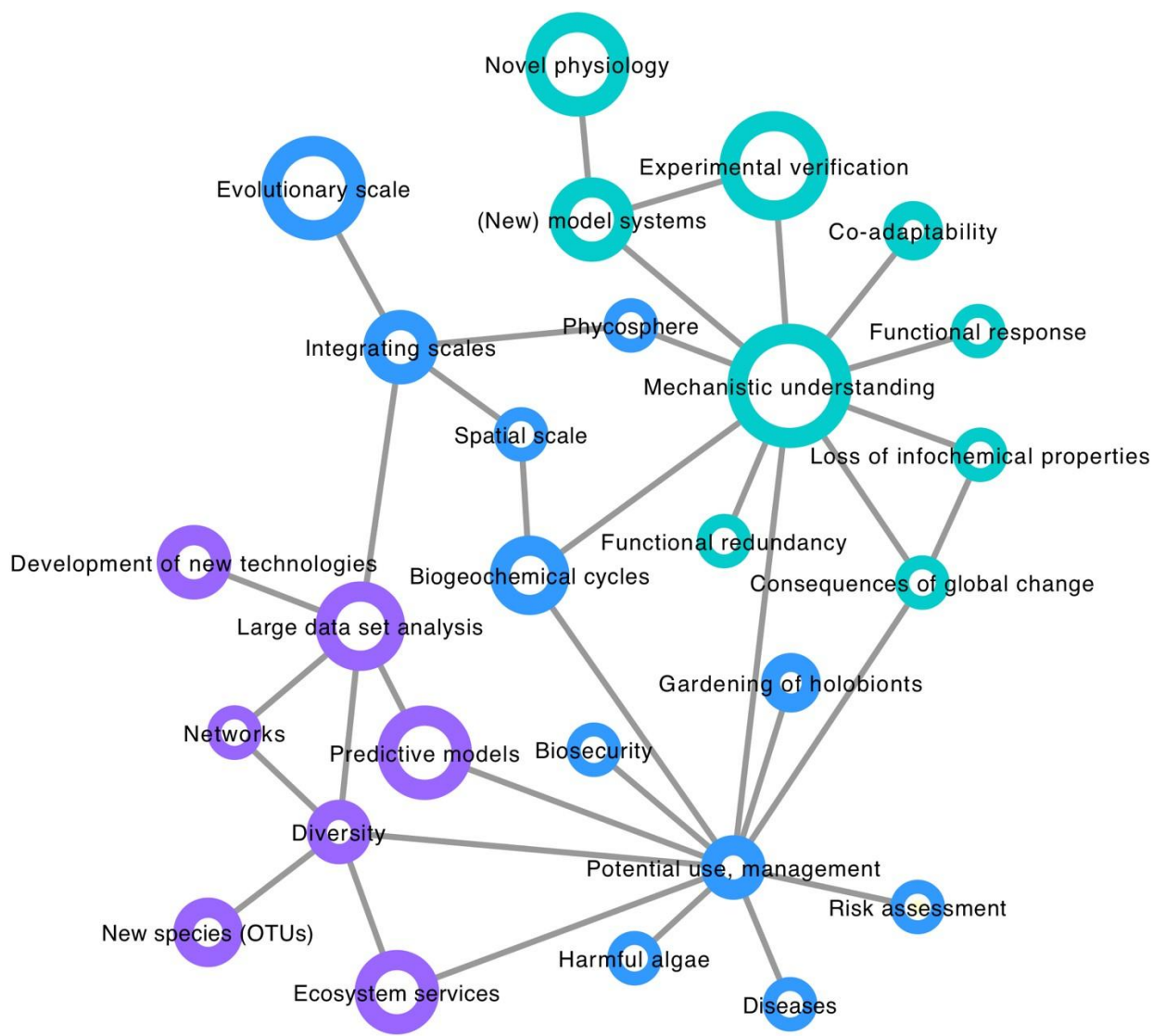

**Figure 3**: Mind map of key concepts, techniques, and challenges related to marine holobionts. This map was generated during the Holomarine workshop held in Roscoff in 2018 (https://www.euromarinenetwork.eu/activities/HoloMarine). The size of the nodes reflects the number of votes each keyword received from the participants of the workshop (total of 120 votes from 30 participants). The two main clusters corresponding to predictive modeling and mechanistic modeling, are displayed in purple and turquoise, respectively. Among the intermediate nodes linking these disciplines (blue) "potential use, management" was the most connected.